\documentclass[twocolumn,           
               showpacs,            
               preprintnumbers,     
               aps,                 
               letterpaper,             
               superscriptaddress,      
               nofootinbib,         
               tightenlines,        
               floats,floatfix      
               ,usenatbib, 11pt, english t
               ]{revtex4-1}
\renewcommand{\thesection}{\arabic{section}}
\renewcommand{\thesubsection}{\thesection.\arabic{subsection}}
\renewcommand{\thesubsubsection}{\thesubsection.\arabic{subsubsection}}

\usepackage{adjustbox}
\usepackage{graphicx}
\usepackage{booktabs, makecell}
\usepackage{lipsum}
\usepackage[english]{babel}
\usepackage{subfigure}
\usepackage{hyperref}
\hypersetup{
    colorlinks=true,
    linkcolor=blue,
    filecolor=magenta,
    urlcolor=cyan,
}

\usepackage{dcolumn}
\usepackage{bm}
\usepackage{listings}
\usepackage{floatrow}
\newfloatcommand{capbtabbox}{table}[][\FBwidth]

\usepackage{blindtext}
\usepackage{graphicx}  
\usepackage{dcolumn}   
\usepackage{bm}        
\usepackage{amsmath,amssymb}
\usepackage{hyperref}
\usepackage{color}
\definecolor{purple}{rgb}{0.58,0.0,0.83}
\usepackage{caption}
\usepackage[toc,page]{appendix}
\usepackage[utf8]{inputenc}
\usepackage{tabularx}
\AtBeginDocument{\selectlanguage{english}}

\newcommand{\jch}[1]{\textcolor{black}{#1}}

\begin{document}
\title{Classification algorithms applied to structure formation simulations}
\author{Jazhiel Chac\'on}
\email{jchacon@icf.unam.mx}
\affiliation{Instituto de Ciencias F\'isicas, Universidad Nacional Aut\'onoma
de M\'exico, Apdo. Postal 48-3, 62251 Cuernavaca, Morelos, M\'exico.}
\author{J.~Alberto V\'azquez}
 \email{javazquez@icf.unam.mx}
\affiliation{Instituto de Ciencias F\'isicas, Universidad Nacional Aut\'onoma
de M\'exico, Apdo. Postal 48-3, 62251 Cuernavaca, Morelos, M\'exico.}
\author{Erick Almaraz}
\affiliation{Instituto de Ciencias F\'isicas, Universidad Nacional Aut\'onoma
de M\'exico, Apdo. Postal 48-3, 62251 Cuernavaca, Morelos, M\'exico.}
\date{\today}

\begin{abstract}

\subsection*{Abstract}
\noindent
 Throughout cosmological simulations, the properties of the matter density field
 in the initial conditions have a decisive impact on the features of the structures formed today.
 In this paper we use a random-forest classification algorithm to infer whether or not dark matter particles, traced back to the initial conditions, would end up in dark matter halos whose masses are above some threshold.
 This problem might be posed as a binary classification task, where the initial conditions of the matter density field are mapped into classification labels provided by a halo finder program. Our results
 show that random forests are effective tools to predict the output of cosmological simulations without running the full process. These techniques might be used in the future to decrease the computational time and to explore more efficiently the effect of different dark matter/dark energy candidates on the formation of cosmological structures.

\textbf{\textit{Keywords}}---Numerical Simulations, $N$-body systems, Machine Learning
\end{abstract}

\maketitle

\section{Introduction}

The evidence gathered over the past twenty years consistently points out that the Universe is made up of about 96\% of dark energy and dark matter. This conclusion is one of the main foundations of the standard cosmological model ($\Lambda$CDM), which despite its success in describing the observations
does not provide yet a complete answer on the physical nature of these components. In this regard, the process of structure formation  is a very useful tool to characterize the properties of the dark sector and to assess
its impact on the historical evolution of the Universe.

Cosmological structure formation is a process determined mainly by the gravitational interaction of dark matter. This process can be broken down into three major stages:

\textsl{Linear regime.}
Initially, dark-matter perturbation modes remain frozen, and they start growing once they enter the causal horizon of the Universe. During this stage, density fluctuations remain small enough to be described by linear perturbation theory.

\textsl{Intermediate regime.} As density fluctuations keep growing, a transition to non-linear regime takes place in which perturbations collapse into denser regions called halos.
This transition process can be described in its essentials
by semi-analytic models, such as the spherical collapse model.

\textsl{Non-linear regime.} Finally, halos group into larger structures that give rise to a cosmic network of filaments and knots. These structures serve as gravitational wells around which visible matter
accretes, so by mapping the distribution of clusters of galaxies, quasars, and gas clouds, we expect to reconstruct the underlying
skeleton of dark matter.

While the linear evolution and the transition to the non-linear regime can be approached by analytical methods, structure formation in the non-linear regime can only be studied using numerical simulations. These simulations  are virtual laboratories by which is possible to study in detail the characteristics of the structures that stem from the dynamics of different candidates of dark matter and dark energy. By comparing the predictions of each model with observations, it is possible to evaluate the feasibility of each one of these scenarios.

Until a few decades ago, numerical simulations were prohibitive
in terms of the amount of the computational resources required, but recently the progress in hardware and the development of new algorithms have made these tools accessible to more research groups.
Still, the cost remain high and for some tasks they will remain out of reach in the foreseeable future. In this sense, there is an incentive to develop artificial intelligence/machine learning solutions that allow the prediction of important features in numerical simulations without the need of completely executing them or,
at most, running a small number of simulations.

The method described in this work is based on a similar approach made in \cite{10.1093/mnras/sty1719}, which uses the ability of machine learning algorithms to
learn complex relationships in large data sets. We aim to find out how much information provide the features of the initial conditions to determine the formation of dark matter halos in cosmological simulations.

The content of this paper is as follows. In section 2 we provide a short review of machine learning fundamentals.
We extend the discussion to two supervised learning algorithms: decision trees and random forests. 
In section 3 we review some metrics of performance.
In section 4 we discuss the problem of halo formation as a binary classification problem. Then, we present the setup of our simulations, the construction of the training and test sets, and the process of hyperparameter tuning.
Finally, in section 7 we present our conclusions and a brief discussion of the results achieved.

%

\section{Machine Learning}

Machine Learning (ML) refers to the set of methods
used to train computers in order to find patterns in data and make inferences without human intervention.
Although this is still a remote possibility for certain applications, currently ML methods are successfully applied to several problems that require the analysis of large volumes of information, for which the cost in terms of human resources needed may represent a serious issue.
ML methods are usually classified into two broad categories:

\textbf{Supervised learning.}
Supervised methods need a set of labeled samples characterized by some features. In this case computers are trained to find a relationship among the features and the labels, so that the labels of new samples may be predicted based on the learned relation. Examples of these algorithms are: Logistic Regression, Decision Trees, Neural Networks, Support Vector Machines, etc (Gron, A. 2017, \cite{10.5555/3153997}). Some works applying these methods in cosmology can be found in \cite{Gomez-Vargas:2021zyl,2013ApJ...772..147X, 2015JCAP...01..038H, 2020arXiv200512276M, Kamdar_2016, Buncher_2020, perraudin2019cosmological}.

\textbf{Unsupervised learning.}
Unsupervised methods are aimed for making inferences on data sets where samples are not labeled. In this case computers are trained to find hidden patterns in the data, letting the information speak for itself. Examples of these algorithms are: Cluster Analysis, Correlation and Principal Component Analysis (PCA) (Goodfellow, I., Bengio, Y., Courville, A., 2016, \cite{10.5555/3086952}). Some applications in cosmology can also be found in \cite{2020arXiv200401393S, 2018MNRAS.478.3410A, D_Addona_2021, Cheng_2020, Geach_2011, Hocking_2017, Cheng_2021}.

\subsection{Supervised Learning}
Since the main goal of this paper is classification we will make use of supervised learning, whose task is the following:
\\
Given a \textbf{training set} of $N$ input-output pairs
\begin{equation}
    (x_{1}, y_{1}), (x_{2}, y_{2}),...,(x_{N},y_{N}),\label{eqn:1.1}
\end{equation}
where each $y_{j}$ is computed using a $y = f(x)$ function, the objective is to find a function $ g $ that approximates the true function $ f $. The function $ g $ is a hypothesis and the variables $ x $ and $ y $ can take any value and not necessarily a numeric one, that is, it can be an attribute. The learning is carried out through a search, within the space of possible hypotheses, of a function that has a good performance, even when it is fed with new examples beyond the training set.
In order to test out  the accuracy of the hypothesis, a \textbf{test set}, different from the training set, is provided. The hypothesis $ g $ is said to generalize well to the function $ f $ if it correctly predicts the values of $ y $ for several inputs.
Furthermore, the dependent variable $y$ can turn out to be categorical, also called qualitative.
The values of a categorical variable are mutually exclusive and in that case the learning problem is called a \textbf {classification} problem, which in turn is referred as Boolean or binary classification if only two values are possible.
\subsection{Decision Trees}
These type of algorithms resemble a chart flow for data, where terminal blocks represent classification decisions. The elements of a decision tree are the root (where the data is stored), the branches (the path the tree takes to make decisions) and nodes (consisting of sets of elements that have a determined characteristic after a decision is made). Given a dataset, we can calculate the inconsistency within the set, or in other words, find its entropy in order to divide or split the set until all data is within a given class (Quinlan R. L., 1986 \cite{quinlan1986induction}).

A decision tree reaches a conclusion by carrying a series of tests. Its nodes perform tests over the attributes on the input values $A_{i}$ and the branches that come from the node are labeled with the possible values of the attribute  $A_{i} = v_{ik}$. The leaf nodes in the tree specify a value that needs to be computed by the function.
A good decision algorithm is developed by splitting the data, so the attribute with the greatest weight or with the highest information gain is obtained, so it is expected to have a correct classification with the least possible number of tests \cite{2014arXiv1407.7502L}.
%

\subsection{Random Forest}
\begin{figure}
    \centering
    \includegraphics[width = \textwidth]{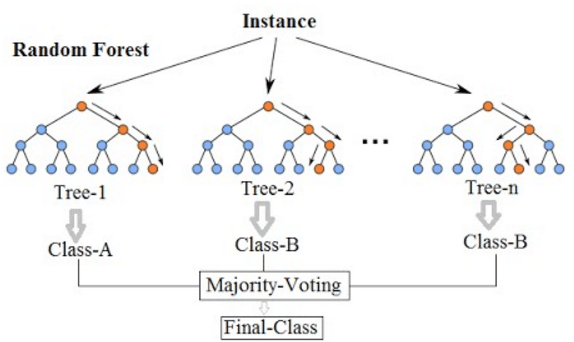}
    \caption{\footnotesize{Sketch of a Random Forest Algorithm. Being an assembly of decision trees, it allows different tests to be carried out on a random selection of attributes, the final class being a vote on a majority obtained in each individual tree. Figure from \href{https://medium.com/swlh/random-forest-classification-and-its-implementation-d5d840dbead0}{Medium}.}}
    \label{fig:3.5}
\end{figure}

The Random Forest algorithm consists of a large number of decision trees that operate together as an ensemble (Hastie, T.; Tibshirani, R.; and Friedman. J., 2009 \cite{hastie_2009}). The ``randomness'' of the algorithm comes from the fact that operations and predictions from the forest are not hierarchically taken, but a subset of elements (like number of trees, number of attributes, length of data, etc.) is taken in a random way. Each individual tree in the Random Forest chooses a class prediction and the class with the most votes becomes the model prediction. This is due to a simple but powerful concept: \textit{The wisdom of the crowds}. The reason that Random Forest is such a good algorithm is because a large number of relatively uncorrelated trees operating together will perform better than any individual model that constitutes it (Breiman, L., 2001 \cite{breiman_random_2001}).
The key is the low correlation among models.
Uncorrelated models can produce joint predictions that are more accurate than any of the individual predictions that make them up. The trees protect each other from their individual errors (as long as those errors are not in the same direction). If some trees have errors, others may get correct predictions, so that as a group the trees can move to the correct direction (see figure \ref{fig:3.5}).

\subsection{Information and Entropy}
%
Decision algorithms like decision trees and random forest perform data division, also called \textit{split} in order to obtain more information after the division is made. This split can be thought of as a way to organize data, thus the learning process should be focused on obtaining a better vision of the analysis process.
This comes directly from information theory: the most valuable information comes from unlikely events rather than events that occur frequently. A way to determine the sought information in a more formal and specific way is by considering that the most probable events provide few information, while the least probable events provide the highest amount of information.

\begin{figure}[t]
    \centering
    \includegraphics[trim = 30mm  0mm 0mm 0mm, clip, width=10.5cm,height=4cm]{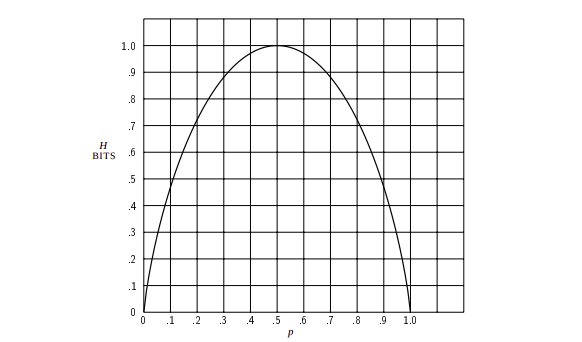}
    \caption{\footnotesize{The entropy for two classes with probability $ p $ and ($ 1 - p $). Shannon's entropy is a way of measuring the relative quantity between the two classes. The entropy value is maximum if there are the same number of classes. Taken from Shannon, C. E. 1948.}}
    \label{fig:Shannon}
\end{figure}

The equation that satisfies these conditions is the information content of an event $x_{i}$
\begin{equation}
    I(x_{i}) = -\log_{2}P(x_{i}),\label{eqn:3.8}
\end{equation}
%
where $P(x)$ is the probability that event $x_i$ occurs.
To account for the whole set of events, a probability distribution is built by using the \textbf{Shannon entropy} (Shannon, C. E., 1948 \cite{doi:10.1002/j.1538-7305.1948.tb01338.x})
\begin{equation}
    H(x) =  -\sum_{i}P(x_{i})\log_{2}P(x_{i}),\label{eqn3.9}
\end{equation}
where the sum of $i$ is over all possible events. That is, the Shannon entropy is the expected amount of information in an event of a probability distribution as observed in figure \ref{fig:Shannon}. The change in the information obtained before and after the division is known as \textbf {information gain}. Therefore, the split is made when the information gain is greater.

\section{Evaluation of classification models}
Evaluating the performance of an algorithm is a fundamental aspect in machine learning. The model must be trained with the \textbf {training set} and then evaluated with the \textbf {test or validation set}, consisting of totally new data not yet evaluated by the algorithm. During the evaluation is important to measure and understand the quality of the classifier and to tune the parameters in the iterative process of discovering the data.
\subsection{ROC curves}

\begin{figure}
    \centering
    \includegraphics[width = \textwidth]{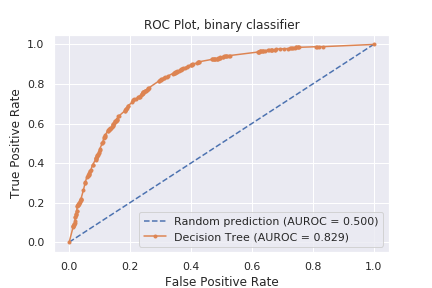}
    \caption{\footnotesize{The ROC curve and the value of the Area Under the Curve (AUC) of a binary classifier. Being a graphic representation, the performance can be evaluated at various prediction thresholds. For different classifiers, the shape of the ROC curve can be very similar, the fairest way to compare them is through the value of the AUC. \href{https://github.com/ChJazhiel/ML_ICF/blob/master/DT_data_nbody.ipynb}{GitHub ChJazhiel}.}}
    \label{fig:3.7}
\end{figure}

Binary classification models are evaluated in the Receiver Operator Characteristics (ROC) space  (Fawcett, T., 2006 \cite{FawcettROC}). A ROC graph is used as a visual representation of the classifier based on its performance. This type of curve shows how the number of correctly classified as true examples changes with respect to the number of incorrectly classified as negative examples.
In ROC space we can define the True Positive Rate (TPR) as
\begin{equation}
    TPR = \frac{TP}{TP + FN},
\end{equation}
and the False Positive Rate (FPR) as
\begin{equation}
    FPR = \frac{FP}{FP + TN}.
\end{equation}
These two quantities are plotted in order to obtain the ROC curve, see figure \ref{fig:3.7} for reference. The FPR measures the fraction of negative examples incorrectly classified as positive, while the TPR measures the fraction of positive examples correctly classified. The convex part of a family of ROC curves can include points located further toward the northwestern boundary of the ROC space. If a line passes through the convex part, then there is no other line with the same slope that passes through another point with a larger TP intersection. In this way, the classifier at that point is optimal under any distribution assumption with that slope (Rokach, L. \& Maimon O. Z., 2008 \cite{rokach2008data}).

\subsection{Area Under Curve (AUC)}

Using continuous-type measures such as ROC curves sometimes can lead to misinterpreting the results. In the case of the ROC curves, for example, for two classifiers there may be an overlap in the curves within the ROC space, so that it becomes difficult to determine which model performed better. If there is no dominant model, it cannot be determined which of them is the best.

The Area Under Curve (AUC) is a quite useful metric to visualize the performance of a classifier, since it is independent of the decision criteria and prior probabilities. Given two classifiers, if the ROC curves intersect then the AUC is an average of the comparison between both models. The AUC does not depend on any imbalance of the training data, so comparing this quantity for two classifiers is fairer and more informative than comparing their misclassification rates, for example. We evaluate the performance of an algorithm with this metric with the range values between 0.5 and 1.0. A value of 0.5 is only as good as a random classifier. Then 0.6--0.7 is considered as a regular classification, 0.71--0.8 a good classification, 0.81--0.9 very good classification and 0.91--1.0 an excellent one.
\subsection{Overfitting and generalization}
Overfitting is a general phenomenon and occurs in all kinds of learning algorithms, even when the target function is not random at all. It becomes more likely as the hypothesis space and the number of input attributes grow, and is less likely as the number of training examples increases.

For random forest and decision tree algorithms, there exist a technique, called \textit{pruning}, that aims to avoid overfitting. Pruning works by removing nodes that are not clearly relevant. The question is, how do you detect that a node is testing an irrelevant attribute? Assuming that a node consists of $ p $ positive examples and $ n $ negative examples. If the attribute is irrelevant,
it would be expected to divide the examples into subsets, so that each one has approximately the same proportion of correctly classified (positive) examples as the complete set, $ p / (p + n) $, in this way the information gain would be close to zero.

How big must the information gain be so that it can be divided on a particular attribute? This question is answered using a significant statistical test where the null hypothesis is that no relationship or underlying pattern exists.
The actual data is then analyzed to calculate the degree to which it deviates from a perfect absence of a pattern. Given a training set $ S $ with input attributes $ A = {a_{1}, a_{2}, ..., a_{n}} $ and a nominal target attribute of unknown fixed distribution $ D $ on the instance space, the goal is to induce an optimal classifier with minimal generalization error.

In other words, given a training set with a finite number of attributes and a set of classes to be determined, find the algorithm that best generalizes the model with a minimal error.

\subsection{Learning curve}
These curves are graphs of the learning performance of the model over experience or time. They are a diagnostic tool widely used in machine learning for algorithms that learn incrementally from a training set. The model can be evaluated on the training set and on a validation dataset after each update during the training. Graphs of the measured performance can be viewed to show the learning curves.

\begin{figure}
    \centering
    \includegraphics[width = \textwidth, scale = 0.7]{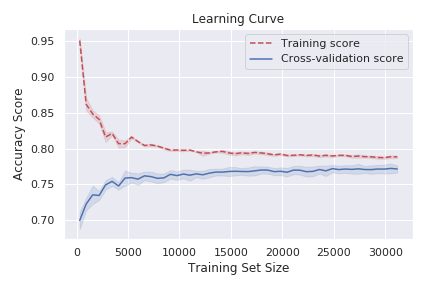}
    \caption{\footnotesize{Learning curves with accuracy metrics of a binary classifier. The algorithm is said to be learning when the validation set curve is close to the training set curve. In this graph, it is observed that the model does not require changing its hyperparameters since the learning set is quite close to the training curve and does not seem to be overfitted. \href{https://github.com/ChJazhiel/ML_ICF/blob/master/DT_data_nbody.ipynb}{GitHub ChJazhiel}.}}
    \label{fig:3.8}
\end{figure}

Reviewing the learning curves of the models during training can be used to diagnose learning problems, such as an underfitting or overfitting model, as well as whether the training and validation data sets are adequately representative, as seen in figure \ref{fig:3.8}.

The evaluation in the validation set offers an idea of how capable the model is of ``generalizing''.
In the space of the learning curve there are two curves:
\begin{itemize}
    \item Train Learning Curve: computed from the training set that gives an idea of how well the model is learning.
    \item Validation Learning Curve: computed from the validation set that gives an idea of how good the model is at generalizing.
\end{itemize}
Because the metrics to evaluate an algorithm are diverse, a simple way to create a learning curve is through accuracy, although it can also be created through an error percentage. To ensure optimal learning, the dataset is divided into subsets of samples called \textbf {$ k $-Fold Cross-validation}. The procedure has a parameter $ k $ that refers to the number of groups the dataset will be divided. It is a simple method to understand and to help the model to decrease variance and avoid bias.
This method has the following steps:
\begin{enumerate}
    \item Shuffle the dataset randomly.
    \item Split the dataset into $k$ groups.
    \item For each unique group:
    \begin{enumerate}
        \item Take a group as a test set.
        \item Use the remaining $k-1$ groups as training set.
        \item Adjust the model with the training set and evaluate it with the test set.
        \item Save the score of evaluation and discard the model.
    \end{enumerate}
    \item Gather model skills using the model score sample.
\end{enumerate}
This approach involves randomly dividing the set of observations into $ k $ groups, of approximately the same size. The first group is treated as a validation set and the method fits the remaining $ k - 1 $ groups (James, G. et al., 2014 \cite{10.5555/2517747}).

\begin{figure}
    \centering
    \includegraphics[width = \textwidth, scale = 0.3]{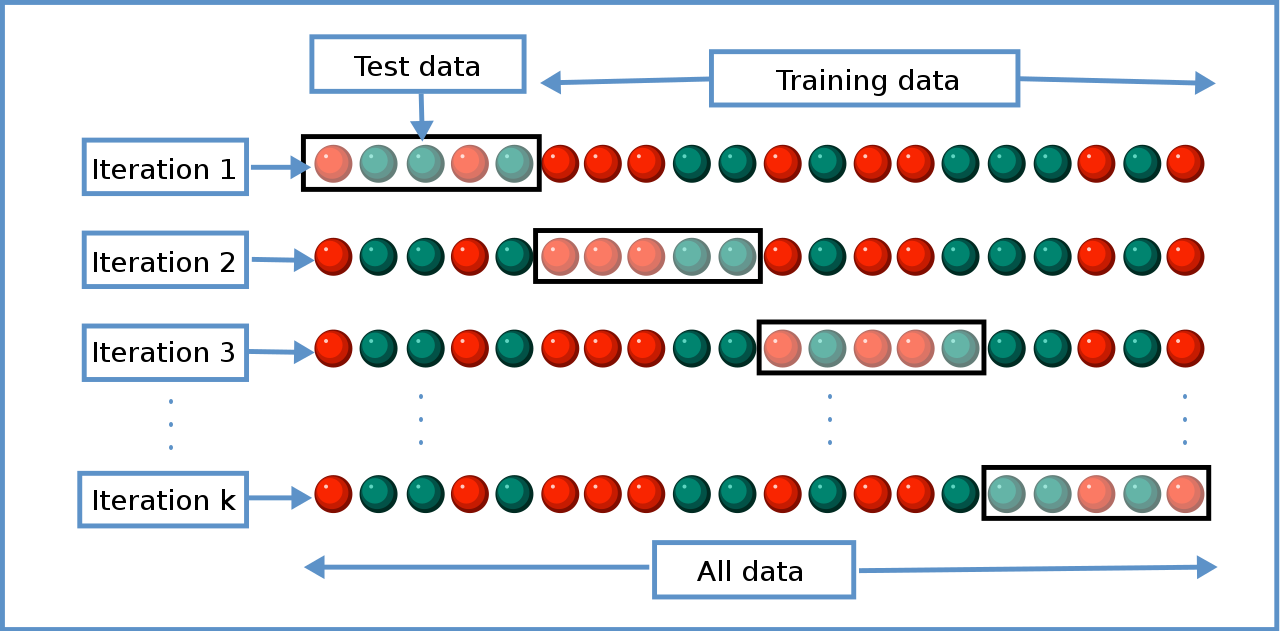}
    \caption{\footnotesize{The $ k $-fold cross validation. The dataset is randomly mixed and a test group is chosen, leaving the rest of the data as training. The iterations are used to carry out this method in a defined way in order to minimize the variance and the bias of the model. (Seni, G.; Elder, J. \cite{Seni-G-2010}).}}
    \label{fig:3.9}
\end{figure}

Graphically, it can be understood with figure \ref{fig:3.9}. From here it is clearly observed how this method mixes and divides the set randomly, so that a small group is the test or validation set and the rest of the data is the training set. This process is carried out recursively to avoid some type of bias or variance of the model.

\section{Numerical Cosmology as a binary classification problem}
After carrying out a series of simulations for various configurations [for more details see (Chac\'on, J.; V\'azquez, J. A.; Gabbasov, R. 2020, \cite{article-Chacon})] we obtain a one-to-one relationship of dark matter halos formed at time $ z = 0 $ with the initial conditions. This allows to identify the dark matter halos, called parents or hosts and in turn we are able to determine substructures or subhalos of the same host. We select a dark matter mass threshold to identify halos, in this way it is possible to identify the particles that end up in a dark matter halo given the mass threshold, as well as those that do not end in a halo, that is, they are free particles or they belong to halos of lower mass. As it can be deduced, this leads to treat the process of dark matter evolution as a classification problem.
%

\subsection{Data selection}
To carry out the process, we chose a cosmological simulation of a  $ \Lambda$CDM Universe made with the cosmological code GADGET-2 (Springel, V. 2005 \cite{2005MNRAS.364.1105S}), with cosmological parameters $\Omega_{m}$ = 0.268, $\Omega_{\Lambda}$ = 0.683, $\Omega_{b}$ = 0.049, $h$ = 0.7. The simulation has a gravitational softening of $ \epsilon $ = 0.89 kpc and it evolves a total of $ 192^{3} $ particles, each with a mass of 1.3 $\times 10^{9} $ M$_{\odot} $ in a box of comoving length $ L = 50 h^{- 1} $ Mpc from $ z = 23 $ to $ z = 0 $. Halos (both host and subhalos) are identified with ROCKSTAR halo finder (Behroozi, P. S., Wechsler,  R. H.\&  Wu, H.-Y., 2013 \cite{2013ApJ...762..109B}). We made two classes, [\textit{Not in Halo}, \textit{In Halo}] by selecting a mass threshold of $ M $ $ \geq $ 1.2 $ \times 10^{12} \, \textup{M}_{\odot} $, so that the class {\textit{in Halo}} will be in halos that exceed this threshold while the particles {\textit{Not in Halo}} are in halos with mass less than said threshold or they are not linked to any halo. The final snapshot counted a total of 4000 dark matter halos whose masses fall within the range $(10^{11} \leq M / M_{\odot} \leq 10^{14})$.

Each particle will have a 10 component vector associated with it and a label: 1 for the class \textit{In halo}, 0 for the class \textit{Not in halo}. The properties of the particles are extracted from the initial conditions of the simulation ($ z = 23 $) and are used as an input data for the decision tree and random forest algorithms. The components are the mass densities centered in each particle linked to the local density of the initial redshift. A subset of all the particles was chosen within the simulation with their respective label. The training was carried out with an 80/20 split of the subset (80 \% training and 20 \% test/validation).

Supervised machine learning algorithms require the use of characteristics of a structured database, in this case there is a structured data set with attributes extracted from the density field. This assignment comes from analytical works related to the halo mass function (HMF) by Press-Schechter (Press, W. H., Schechter, P. 1974 \cite{1974ApJ...187..425P}). This function predicts the density of the number of halos of dark matter depending on their mass and the density field. The density will form a halo of a certain mass $ M $ at a redshift $ z $. If it exceeds a critical value $ \delta_ {c} (z) $, these values will be called overdensities at a given redshift $ z $.

The main idea is that the matter of a halo will be enclosed in a dense spherical region, where the density contrast  will be given by the relation
\begin{equation}
    \delta(\textbf{x}) = \frac{\rho(\textbf{x}) - \Bar{\rho}}{\Bar{\rho}},\label{eqn:4.1}
\end{equation}
where $\Bar{\rho}$ is the average matter density of the Universe. For a sphere of radius $R$ (Dodelson, S., 2003 \cite{dodelson2003modern}), it is well understood that the overdensity is
\begin{equation}
    \delta(\textbf{x},R) \equiv \int d^{3}\textbf{x}'\delta(\textbf{x}')W_{R}(\textbf{x} - \textbf{x}').\label{eqn:4.2}
\end{equation}
In equation (\ref{eqn:4.2}), $W_{R}$ is a window function of the \textit{top hat} model, given by
\begin{equation}
W_{R} = \left\lbrace
\begin{array}{ll}
\frac{3}{4\pi R^{3}} & \textup{if} \; |\textbf{x}| \leq R\\
0 & \textup{if} \; |\textbf{x}| > R.
\end{array}
\right.\label{eqn:4.3}
\end{equation}
A window function with radius $ R $ corresponds to a mass scale $M = \Bar{\rho} V(R)$. The expected value of the overdensity in equation (\ref{eqn:4.2}) is the normalization term of the power spectrum $ \sigma_{R} $
\begin{equation}
    \sigma^{2}_{R} = \langle \delta^{2}(\textbf{x}, R)\rangle.\label{eqn:4.4}
\end{equation}
The choice of attributes of the structured data for the machine learning algorithms reside in the density contrasts calculated with the top hat type window function that is derived from a mass scale $M_{R}$ in the radius $R$, centered on the position of a particle, from the initial conditions and the initial redshift $z = 23$. The result is a quantity of 10 overdensities $\delta_{1},...\delta_{10}$, each one associated with their respective class or label. 
\jch{This 10-component vector along with their corresponding labels makes a dataset that achieves well performance. That is, when we chose more than 10 component vectors, which means using bigger regions of study, we saw no further improvement in our training.  On the other hand, if less vectors were studied we observed a low accuracy performance in the algorithms . The mass range was selected taking into account two main features. First, the data we selected had more massive halos and less massive halos and free particles, the mass range was an average of the halos found. Second, the mass range was also a calculation from the approximation of the spherical collapse model, which gave us the number for the threshold.}

\begin{figure*}[t!]
    \centering
    \includegraphics[width =\textwidth]{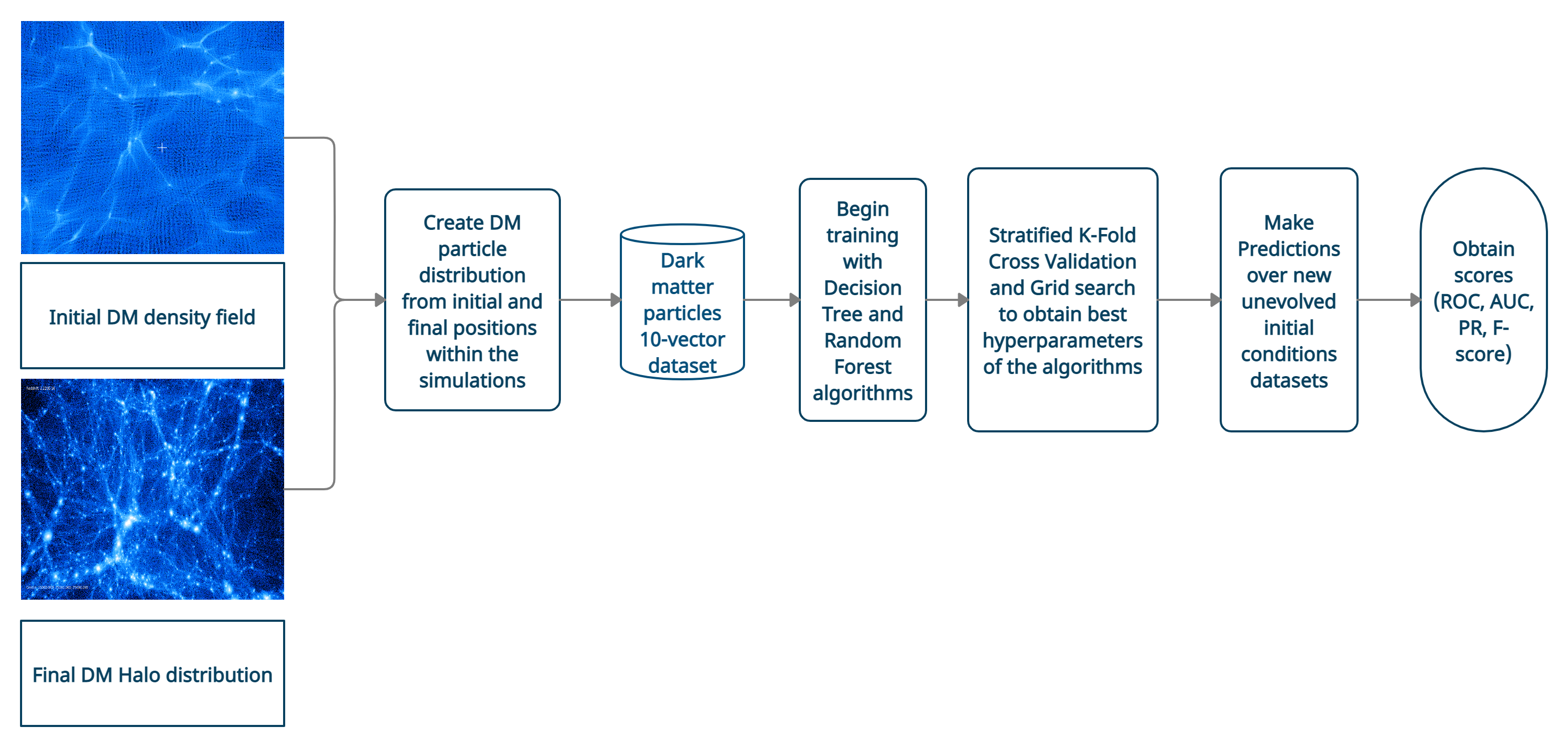}
    \caption{\footnotesize{Diagram of the method to select the properties of the initial   density field conditions that will eventually form the structure in the simulation. The process starts from extracting properties of the initial conditions in the local neighborhood of the density field around dark matter particles and associates them to the final position in the halo distribution. The final classification \textit{Not in halo}, \textit{In Halo} depends on the mass threshold chosen to determine whether a dark matter particle will belong in a halo or if it's not bound to any other object.}}
\end{figure*}

\subsection{Training}\label{seccion:4.2}

Algorithms used for this section were decision trees and random forest, included in the machine learning package from \textbf{Scikit-Learn} (Pedregosa, F. and et al. 2011 \cite{scikit-learn}). The initial number of particles was 50,000 randomly selected, but a preprocessing was performed before i.e. labels [\textit{Not in Halo}, \textit{In Halo}] were converted to a set of labels 0 and 1, respectively.
After this preprocessing, the total number of particles is reduced to 28,600. The algorithms were tested for both quantities and no reduction in performance was observed when reducing the number of particles. The dataset is selected randomly so there is no bias when performing the classification. The training set, as mentioned above, is 80 \% of the total particles, so that 22,880 particles served as the training set, while the validation set was the remaining 5720 particles.

Both decision trees and random forest algorithms were fine tuned by making tests in a hyperparameter grid. This grid had elements such as the maximum depth of the tree, the element split criterion, the maximum number of particles per node, the minimum number of particles to make a split, and in case of random forest, the total number of estimators. Starting from the number of estimators in random forest at 100, increasing by 100, the depth of the tree starting at 1 and reaching 20 increasing by 1, the minimum number of particles at 50, up to 200, increasing by 50, thus finding the optimal values in order to avoid blind testing. The optimal hyperparameters are highlighted in table \ref{Tabla:4.1}, being the same in almost all values except for the number of estimators, exclusive to random forest.
The codes already trained predict the final label of the particles in the test set, which is compared with the real labels in order to obtain the performance of each algorithm. This evaluation was carried out under two tests, the ROC curve along with the AUC of the ROC curve and the learning curve.

\begin{table}[t]%
\caption{Optimal hyperparameters for the algorithms.}
\label{Tabla:4.1}%
\centering
\begin{tabularx}{\linewidth}{@{\extracolsep{\fill}} l c c }
\toprule%
\hline
Description  &Symbol & Value \\
\toprule%
\hline
Decision criteria & \texttt{criterion} & \texttt{entropy}\\
Max depth & \texttt{max\_depth} & \texttt{8}\\
Class balance & \texttt{class\_weight} &\texttt{balanced}\\
No. of estimators & \texttt{n\_estimators} & \texttt{2000}\\
Min. No. of particles & \texttt{n\_particles} & \texttt{200}\\
\hline
\end{tabularx}
\end{table}

\section{Dark matter particles classification}

\begin{figure}[t]
    \centering
    \includegraphics[width = \textwidth, scale = 0.1]{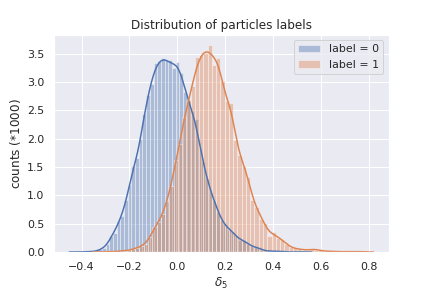}
    \caption{\footnotesize{Histogram of classes for an overdensity obtained in the data preprocessing. The shape of the distribution suggests:
    1) It is not necessary to do a data rescaling, since the similarity with a Gaussian curve is evident.
    2) The use of the ROC curve metric is sufficient due to the distinction of classes in this range of overdensity values.}}
    \label{fig:4.1}
\end{figure}

Due to the probability distribution obtained for each overdensity range, it is not necessary to perform an extensive preprocessing (see figure \ref{fig:4.1}). This figure describes the class distribution (\textit{Not in Halo}: \texttt{label = 0}, \textit{In Halo}: \texttt{label = 1}) depending on the density contrast $\delta_{i}$. The overdensities $\delta_{5}, \; \delta_{6}, \; \delta_{7}$ correspond to mass values $1.2 \times 10^{12} \textup{M}_{\odot}, \; 2\times10^{12}\textup{M}_{\odot}, \; 1.1\times10^{13}\textup{M}_{\odot}$ respectively and radius $R$ ranging from 3 kpc to 6 kpc, coinciding with the limit that we chose to make a decision ($1.2\times10^{12}\textup{M}_{\odot}$). The classification algorithms do not need a rescaling of characteristics since they make decisions through the gain of information, unlike other methods where a subtle difference, for example, the same distance (with different units) can affect the performance of the algorithm. The results are the probability of each class for all particles. That is, the result of belonging to one class or another is determined by a probability threshold value.

After taking this into account, the performance of the algorithms is quantified. A perfect classifier will consist of true positive and true negative values in its confusion matrix. The true positive rate (TPR) and the false positive rate (FPR) are the characteristic quantities of a ROC curve.

\begin{figure}[t]
    \centering
    \includegraphics[trim = 10mm  1mm 10mm 3mm, clip, width=8.cm, height=5.cm]{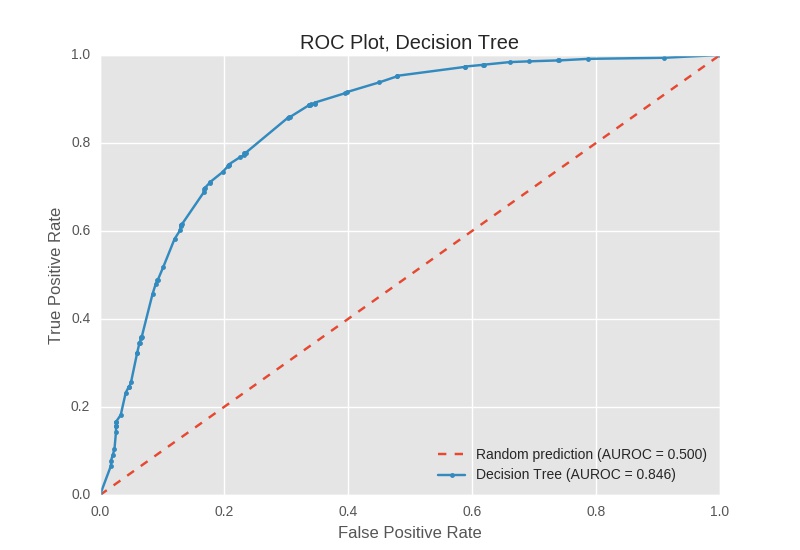}\\
    \includegraphics[trim = 10mm  1mm 10mm 3mm, clip, width=8.cm, height=5.cm]{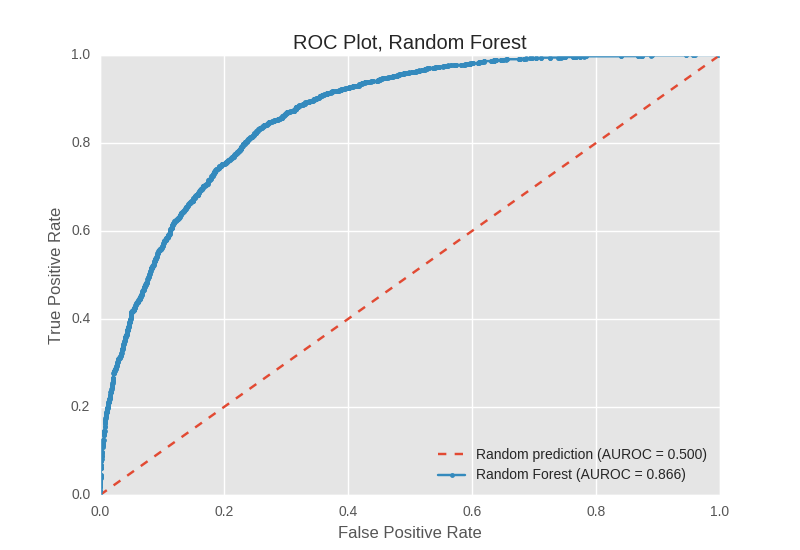}
    \caption{\footnotesize{ROC curves of a decision tree and a random forest algorithm trained in the GADGET simulation. The performance is remarkable given that both have an AUC value $ \geq $ 0.8, highlighting the improvement that random forest has over the decision tree.}}
    \label{fig:4.2}
\end{figure}

The number of particles correctly classified (TPR) and the number of particles incorrectly classified as true (FPR) are shown in figure \ref{fig:4.2}. The tests performed for the decision tree gave an accuracy value of 0.77 $ \pm $ 0.01, with a value of AUC $ = $ 0.846. For the random forest, the accuracy was 0.78 $ \pm $ 0.01 and AUC value $ = $ 0.866.

It can be seen in figure \ref{fig:4.2} that TPR decreases as the FPR also decreases. Decision algorithms have been able to predict in a good way whether a particle will end up in a halo or not, depending on the overdensity of the dark matter density field from the initial conditions.

\begin{figure}[t]
    \centering
    \includegraphics[trim = 1mm  1mm 1mm 3mm, clip, width=8.cm, height=5.cm]{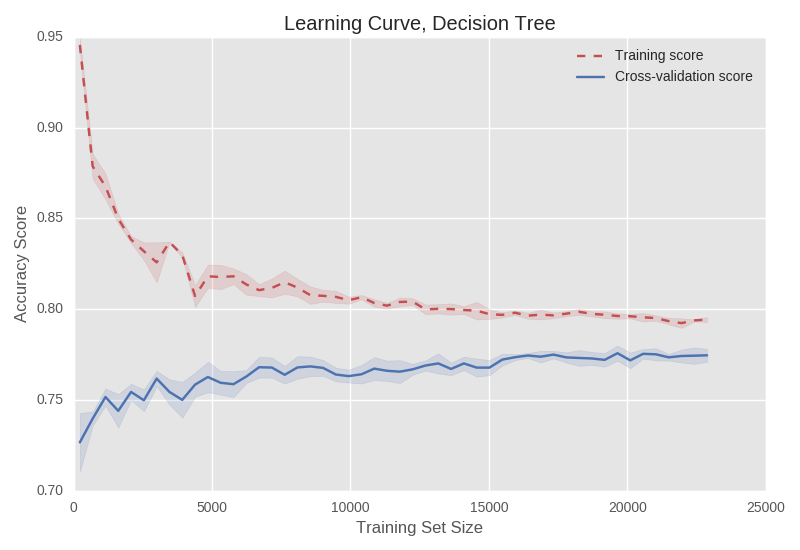}
    \includegraphics[trim = 10mm  1mm 10mm 3mm, clip, width=8.cm, height=5.cm]{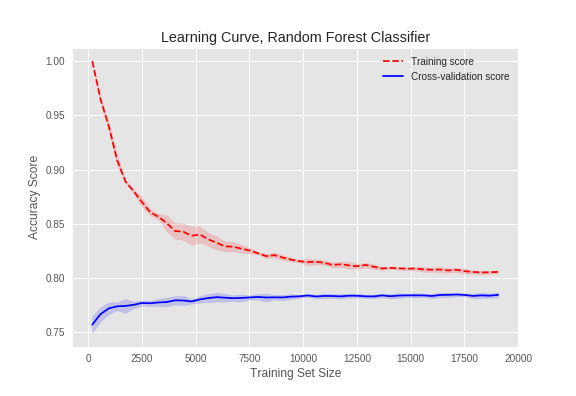}
    \caption{\footnotesize{Learning curves of the decision tree and random forest algorithms. The training curve starts out very high because we have few samples to make a prediction. As the samples increase, the learning curve of the validation set also increases, showing that there is neither overfitting nor underfitting. It is noteworthy that the learning curve of the random forest has less variance, since the low correlation between characteristics prevents a change in this value.}}
    \label{FIG:4.4}
\end{figure}

Also as part of the algorithm evaluation, the learning curves of the decision tree and random forest are described in figure \ref{FIG:4.4}. The upper part corresponds to the decision tree, while the lower part represents the random forest. Both methods adjust their performance well as the number of tests and validation elements increase, reaching a value almost parallel to that reported by the training set. As the training curves neither increase nor the validation curves fall after performing the tests with \textit {cross-validation} it is possible to conclude that both methods are well fitted. 
\jch{Additionally  other methods like Logistic Regression and even Naive Bayes were tested, nevertheless the process of decision making of those algorithms does not quite fit the overall result we aim for. The use of Random Forest and Decision Trees has the advantage of being more visual, less biased and with no overfitting when it comes to making decisions for unseen data.}

\section{Test on new initial conditions}
The training and tests sets used on the classification algorithms have been generated during the $N$-body simulations. The advantage is that an evaluation can be carried out on an independent set of initial conditions and the prediction effectiveness can be tested. For this purpose, four new sets of initial conditions were created, these being listed in Table \ref{Tabla:2}.

\begin{table*}[t]%
\caption{Initial conditions for the cosmological simulations}
\label{tabla:2.2}\centering%
\begin{tabularx}{\textwidth}{@{\extracolsep{\fill}}   l c c }
\hline%
Description & Symbol & Value\\\hline%
Dark Matter Density & $\Omega_{m}$ & 0.268\\
Dark energy Density & $\Omega_{\Lambda}$ & 0.683\\
Baryonic Matter Density & $\Omega_{b}$ &0.049\\\hline
Boxsize & $L$ & 50 Mpc\\
Particle No. & $N$ &  $192^{3}$\\
Initial Redshift & $z_{init}$ & 23\\
Final Redshift & $z_{f}$ & 0\\
Hubble's Parameter & $h$ & 0.7\\
Matter Power Spectrum Normalization & $\sigma_{8}$ & 0.8\\
Seed for IC-generator & \texttt{Seed} & 100,200,300,400\\\hline
\; \; \; \; \; \; \; \; \; \; \; \; \; \; \;\; \; \; \; \; \; \; \; \; \; \; \;  \; \; \;
Another Technical Quantities \\
ErrorTolIntAccuracy & & 0.025 \\
MaxRMSDisplacementFact & &0.2 \\
CourantFact & &0.15\\
MaxSizeTimestep & &0.03\\
ErrorTolTheta & &0.5\\
TypeOfOpeningCriterion & &1\\
ErrTolForceAcc & &0.005\\
\hline \label{Tabla:2}
\end{tabularx}
\end{table*}

In three of them the initial seed was changed, which is essentially a pseudo-random creator of numbers that are transmitted to the positions of the particles into the initial conditions. In another simulation, the gravitational smoothing length parameter $\epsilon$ was also changed (the first simulation has a value of $\epsilon_{1} = 0.89$ kpc). The value of $\epsilon_{1}$ is not trivial, this arises from the analytical calculation of force and acceleration of a top-hat spherical collapse model. 
\jch{ In this calculations, the gravitational softening $\epsilon_{1}$ widely fits how acceleration between particles in a simulation should be in order to acquire results that are in agreement with other cosmological studies. The three different seeds were determined to add stochasticity to the particle generation in the initial conditions, this stochasticity was required in order to make the decision less biased and to prove the generalization of the decision algorithms.}

\begin{figure}[t]
    \centering
    \includegraphics[width = \textwidth]{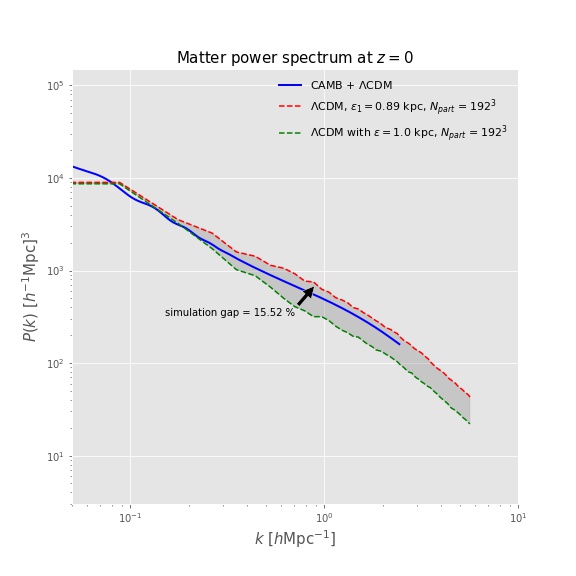}
    \caption{\footnotesize{Matter power spectrum of the new initial conditions ($\epsilon_{2} = 1.0$ kpc) and the  simulation above; the spectrum obtained with CAMB is shown in the solid line \cite{Lewis:2002ah}. The difference between both simulations is indicated in the figure and is approximately 15 \%. The power spectrum was obtained in the same way as in the previous realizations. It is evident that the distribution of matter for the new conditions is different, since there is less structure formation.}}
    \label{fig:4.5}
\end{figure}

In the new simulation we changed the smoothing length, by increasing it to $\epsilon_{2} = 1$ kpc. Remembering that this length is the minimum distance that two dark matter particles can be together in the simulation  (Zhang, T.; Liao, S.; Li, M.; Gao, L., \cite{2019MNRAS.487.1227Z}), the distribution of matter is expected to change, which can be corroborated with the mass power spectrum, observed in figure  \ref{fig:4.5}. The properties around the particles in the dark matter density field were again extracted and a new evaluation of the performance of the decision tree and random forest was carried out.

\begin{figure}[t]
    \centering
    \includegraphics[trim = 10mm  1mm 10mm 3mm, clip, width=8.cm, height=5.cm]{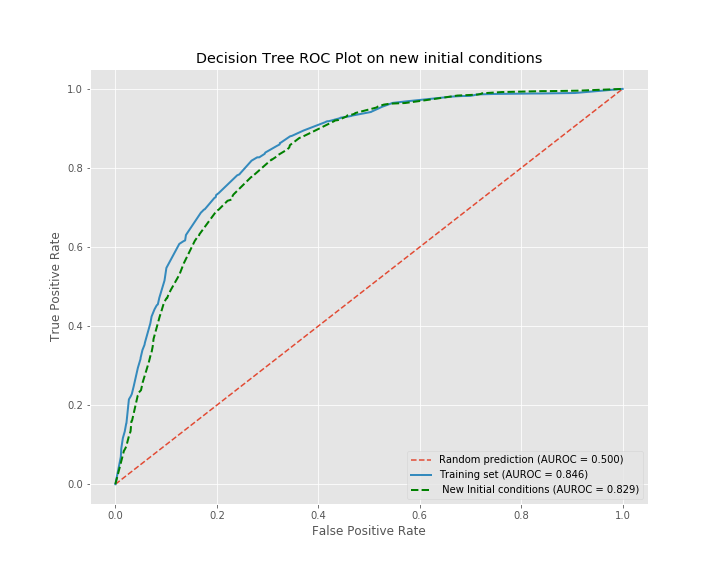}
    \includegraphics[trim = 10mm  1mm 10mm 3mm, clip, width=8.cm, height=5.cm]{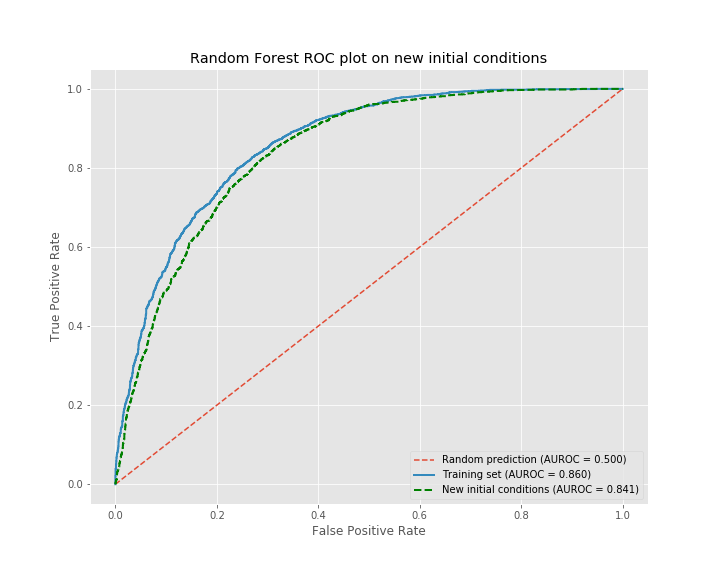}
    \caption{\footnotesize{ROC curves of the decision tree and random forest algorithms of the initial conditions with a new gravitational smoothing $ \epsilon $, compared to the performance previously shown. The curves are fairly consistent. The value of the AUC fell $ \sim $ 2 \%. The tests demonstrate the great capacity of the algorithms to predict the final labels of different simulations.}}
    \label{fig:4.6}
\end{figure}

Figure \ref{fig:4.6} shows the comparative ROC curve of the two algorithms in the realization with the new gravitational smoothing $ \epsilon $, when training and testing them with the data from the initial simulation, as well as when doing the test with the new initial conditions, without carrying out a complete computational run. The upper part shows the performance of the decision tree in the previous training and testing set and the prediction for the new initial conditions. The bottom part shows the same for the random forest. Decision algorithms produce consistent ROC curves for the new set of initial conditions. The AUC on both approaches fell $ \sim $ 2 \% since there was less structure formation.

\begin{figure}[t!]
 \centering
    \includegraphics[trim = 10mm  1mm 10mm 3mm, clip, width=8.cm, height=5.cm]{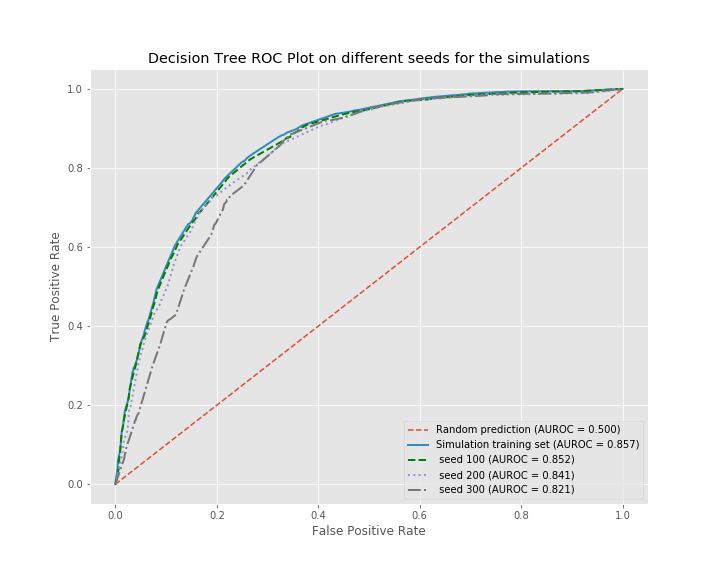}
    \includegraphics[trim = 10mm  1mm 10mm 3mm, clip, width=8.cm, height=5.cm]{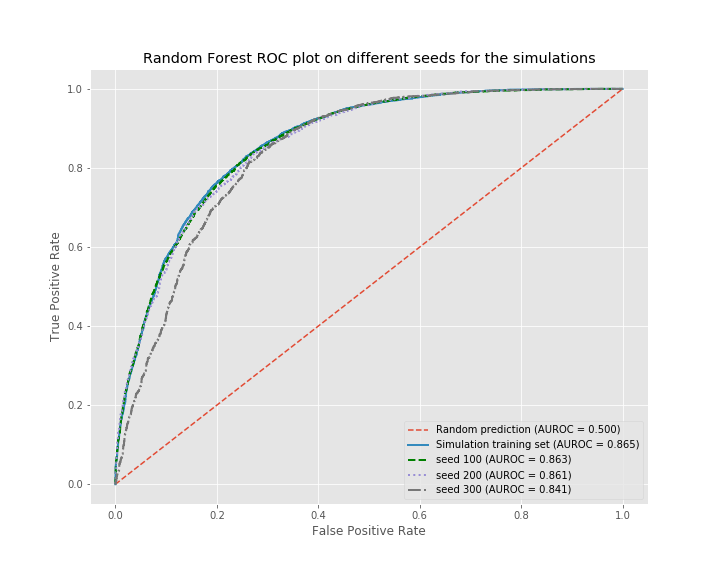}
    \caption{\footnotesize{ROC curves of the decision tree and random forest algorithms of the initial conditions whose seeds were different. The AUC drops an average of 2.2 \% for new realizations. The generalization of the predictive power of the training is evident since the algorithms are able to decide in a good way the final destination of the dark matter particles from their position at an initial moment.}} \label{fig 4.9}
\end{figure}

On the other hand, the change of seed realizations had a performance similar to that described in the previous paragraph. We had the $\Lambda$CDM simulation as a training set, and the performance was tested on the new initial conditions.
Though the complete simulations were not executed, the algorithms  were  able to identify the classification of dark matter particles that fell or not into halos of dark matter given a threshold value.
Figure \ref{fig 4.9} shows the performance of the decision tree in the upper part, and the random forest in the lower part. Both algorithms perform proficiently with their $ \Lambda $CDM simulation training counterparts.
The only change in the initial conditions of these new realizations was in the pseudo-random seed generator, so that the predictive power of the algorithms becomes more evident with this figure.

\begin{figure}[t!]
 \centering
    \includegraphics[trim = 10mm  1mm 10mm 3mm, clip, width=8.cm, height=5.5cm]{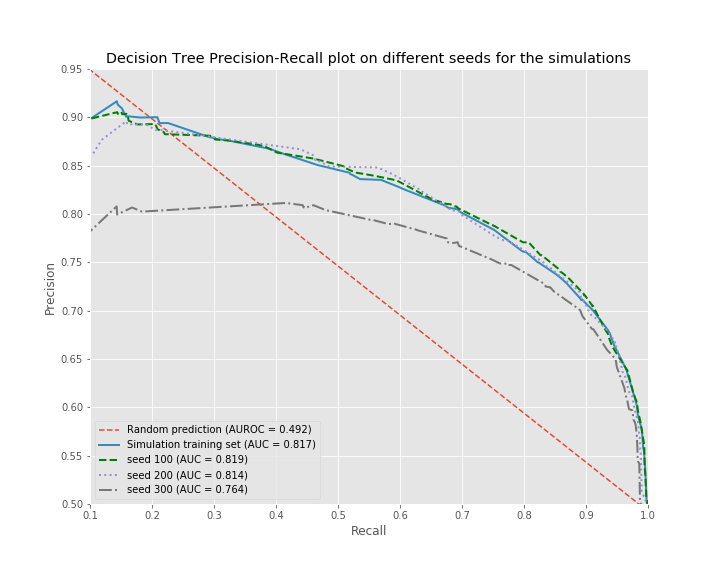}
    \includegraphics[trim = 10mm  1mm 10mm 3mm, clip, width=8.cm, height=5.5cm]{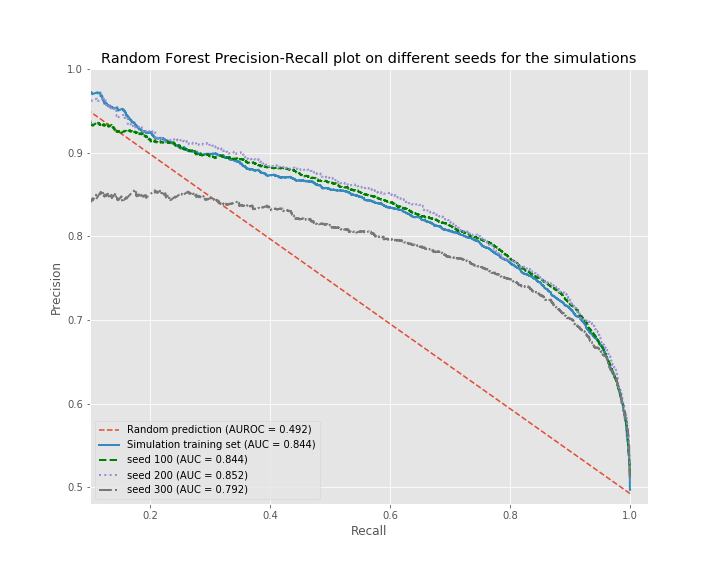}
    \caption{\footnotesize{Precision-Recall scores obtained in the learning process of decision algorithms. We see that for a particular seed, the decision tree Precision drops up to 0.78, this is likely because there were more cases of False Negatives in this dataset, whereas the random forest only drops to 0.85, meaning that the random forest performs generally better than the decision tree. The figure suggests a good accuracy performance on the new seeds generated for the training process, resulting in a reliable algorithm that serves as our binary classifier for dark matter particles within the simulation.}} \label{fig: 13}
\end{figure}

In figure \ref{fig: 13} we can see the Precision-Recall scores obtained for the decision algorithms (decision tree and random forest), with Precision defined as
\begin{equation}
    Precision = \frac{TP}{TP + FP},
\end{equation}
and Recall defined as
\begin{equation}
    Recall = \frac{TP}{TP + FN}.
\end{equation}
We can see that Recall is another name for False Positive Rate. As we know, precision and recall both indicate accuracy of the model. Precision means the percentage of the results which are relevant, while recall refers to the percentage of total relevant results correctly classified by the algorithms. The AUC of the PR curve for both decision tree and random forest are between $[0.76,0.82]$, and $[0.78,0.84]$, respectively. The result given suggests that the learning process did have a good trade-off between precision and recall across all different IC seeds.
We therefore conclude that the overall accuracy had no impact while making a small change in the initial conditions. Additionally, the F1-score, defined as
\begin{equation}
 F_{1} = 2\frac{Precision \cdot Recall}{Precision + Recall},
\end{equation}
and $F_{\beta}$ score defined as
\begin{equation}
F_{\beta} = (1 + \beta^{2}) \frac{Precision \cdot Recall}{\beta^{2}Precision + Recall},
\end{equation}
where $\beta = 0.5$ is chosen such that Recall is considered $\beta$ times as important as Precision, were calculated. Table \ref{Tabla:3} lists the averaged weighted F-scores. This result determines the overall good performance by both algorithms and remarks better results obtained by the random forest.

\begin{table}[t]%
\caption{F-score of decision algorithms.}
\label{Tabla:3}%
\centering
\begin{tabularx}{\linewidth}{@{\extracolsep{\fill}} l c c }
\toprule%
\hline
 Algorithm & $F_{1}$ & $F_{\beta}$  \\
\toprule%
\hline
Decision Tree & 0.775 & 0.777\\
Random Forest & 0.780 & 0.0.781\\
\hline
\end{tabularx}
\end{table}


Labels predicted by the machine learning algorithms are computed from the density properties of the initial conditions. In simulations, the algorithms are able to predict the final classification result with fairly good accuracy. By carrying out a new test for the different initial conditions, without running the full simulation, both methods were able to predict the final label accurately.
Granted a  non negligible change in initial conditions did affect the outcome.


\section{Final discussion and future work}

Throughout the course of this work, there were several factors that brought up our attention regarding our results. One of them being the use of different simulation parameters, which are independent form any physical properties. This led to test several simulations in order to obtain reliable results,  otherwise they could be misleading and would lack of physical interpretation. Similar to many other fields, numerical simulations need a deep understanding of the physical processes behind as well as the parameters of the simulation, which it is key to know in order to obtain reliable results. 


It has been emphasized that the algorithm training process has to be more refined, since the number of selected particles represents a minor contribution of the total number of particles within the simulation. This can certainly be decisive since at the end there was an array of approximately $ 317,680 $ elements that should have been used as training data. Even so, another test was performed for a choice of $ 57,000 $ particles, performing the procedure described in section 4.2. The AUC  of the algorithms used did not show an improvement since both realizations have similar results (0.85 for decision tree and 0.86 for random forest). This fact shows that the use of a larger volume of particles is not decisive in the identification process.

\jch{Finally, The aim of this work was to show how we can use Tree-like decision algorithms to aid cosmological simulations and predict the outcome of a future run without the need of evolving dark matter particles with codes that consume a lot of time, giving results that are similar to the ones obtained in a full run. Furthermore, we saw that using less data, we obtained a good overall result in predictions for new initial conditions datasets, resulting in models that are able to learn the relationship between the initial conditions (position and region of overdensity) of the dark matter particles and their final position within halos given a certain threshold mass.
Additionally, the data used for this work can be retrieved from our GitHub repository\footnote{\href{https://github.com/ChJazhiel/ML_ICF/blob/master/DT_data_nbody.ipynb}{GitHub ChJazhiel}.}, in which the data and code necessary to perform our analysis is available. We encourage the reader to visit this site and perform their own tests.}

\subsection{Numerical simulations assisted with artificial intelligence}

There is another alternative to the complete realization of a numerical simulation, using Generative Adversarial Networks (GAN). These networks basically take databases or images from which the algorithm generates two networks, a generator and a discriminator. The networks begin a competition between each other, both networks were trained with the same data set, but the first must try to create variations of the data that it has already seen. The discriminatory network must identify whether the image created is part of the original training or is a false image that the generative network created.
The more datasets generated the better the generative network is at creating them, and the more difficult it is for the discriminatory network to identify whether the image is real or false. The generative network needs the discriminator to know how to create an imitation so realistic that the second one cannot distinguish it from a real image.

In this regard, numerical simulations come in handy, because the displacement density field is shown as a 3-dimensional image with 3 channels, each channel corresponds to the displacement vector, the deep learning model takes the displacements of the low-resolution simulation and generates a possible high-resolution realization, so this result can be seen as a high resolution simulation with more particles and higher mass resolution (Li, Y.; Ni, Y.; Croft, R. A. C.; Di Matteo, T.; Bird, S.; Feng, Y., 2020, \cite{2020arXiv201006608L}).
A goal in the future will be understanding and implementing deep learning frameworks that can yield better resolution simulations without requiring more  computational time and resources, and obtaining results similar to the ones obtained in the numerical code method.

\section*{Acknowledgments}
\noindent
J.A.V. acknowledges support from FOSEC SEP-CONACYT Ciencia B\'asica A1-S-21925, FORDECYT-PRONACES-CONACYT 304001 and UNAM-DGAPA-PAPIIT IA104221.

\noindent
J. Ch. would like to thank Sebastien Fromentau and Octavio Valenzuela who offered feedback about the process for this work. Additionally, the author acknowledges support from project CONACYT 282569, february-july 2019.

\bibliographystyle{ieeetr}
\bibliography{main}

\end{document}